\documentclass[conference]{IEEEtran}
\usepackage{booktabs} 
\usepackage{graphicx}
\usepackage{balance}
\usepackage{comment}
\usepackage{amsmath}
\usepackage{amsthm}
\usepackage{amsfonts} 
\usepackage{makecell, rotating, multirow, diagbox}
\usepackage{arydshln}
\usepackage{url}
\usepackage{subfigure,algorithmic,lineno,enumerate,epsfig,epstopdf,pifont,setspace,dsfont}
\usepackage[accsupp]{axessibility}
\usepackage{amssymb}
\usepackage{times}
\usepackage{colortbl}
\usepackage[ruled, vlined, linesnumbered]{algorithm2e}%
\usepackage{paralist}
\usepackage[dvipsnames]{xcolor}
\usepackage{bm}
\usepackage{microtype}
\usepackage{pifont}
\usepackage{graphicx}
\usepackage{subfigure}
\usepackage{tablefootnote}
\usepackage{enumitem}
\usepackage{subcaption}
\usepackage{caption}
\usepackage[pagebackref,breaklinks,colorlinks,citecolor=green]{hyperref}
\begin{document}

\title{DeepFake-o-meter v2.0: An Open Platform for DeepFake Detection}


\author{%
  \IEEEauthorblockN{%
    Yan Ju$^{1}$,
    Chengzhe Sun$^{1}$,
    Shan Jia$^{1\dagger}$,
    Shuwei Hou$^{1}$,
    Zhaofeng Si$^{1}$,
    Soumyya Kanti Datta$^{1}$,
    \\Lipeng Ke$^{2}$,
    Riky Zhou$^{1}$,
    Anita Nikolich$^{3}$, 
    Siwei Lyu$^{1}$
  }%
  \IEEEauthorblockA{$^1$ University at Buffalo, State University of New York, Buffalo, USA}%
  \IEEEauthorblockA{$^2$ Amazon Lab126, Sunnyvale, California, USA}%
  \IEEEauthorblockA{$^3$ University of Illinois Urbana-Champaign, Illinois, USA}%
}


\maketitle

\begingroup
\renewcommand\thefootnote{$\dagger$}\footnotetext{Corresponding Author (shanjia@buffalo.edu)}
\endgroup

\begin{abstract}
Deepfakes, as AI-generated media, have increasingly threatened media integrity and personal privacy with realistic yet fake digital content. This work introduces an open-source and user-friendly online platform, DeepFake-O-Meter v2.0, that integrates state-of-the-art methods for detecting DeepFake images, videos, and audio. Built upon DeepFake-O-Meter v1.0, we have significantly upgraded and improved the platform architecture design, including user interaction, detector integration, job balancing, and security management. The platform aims to offer everyday users a convenient service for analyzing DeepFake media using multiple state-of-the-art detection algorithms. It ensures secure and private delivery of the analysis results.
Furthermore, it serves as an evaluation and benchmanrking platform for researchers in digital media forensics to compare the performance of multiple algorithms on the same input. We have also conducted a detailed usage analysis based on the collected data to gain deeper insights into our platform's statistics. This involves analyzing four-month trends in user activity and evaluating the processing efficiency of each detector. 

\end{abstract}
\section{Introduction}
\label{sec:intro}


Generative AI models are creating highly realistic synthetic images~\cite{karras2019style, ho2020denoising,rombach2022high}, videos~\cite{MakePixelsDance,cho2024sora,liu2023evalcrafter}, and audio~\cite{huang2023make,liu2023audioldm,kreuk2022audiogen,borsos2023audiolm}, with both improved quality and faster processing time. The rise of generative media poses a significant threat of impersonation and disinformation, commonly known as DeepFakes. They erode the public trust in domains such as social media, politics, military, geospatial intelligence, and cyber-security. 

Correspondingly, a variety of detection methods have been developed to expose the generated image, video, and audio~\cite{lyu2020deepfake,masood2023deepfakes, khanjani2023audio, malik2022deepfake}. Current DeepFake detection methods typically rely on deep neural networks to automatically learn and extract distinctive features from the media. These features are then fed into a binary classifier to determine whether the media is real or fake.
To detect generated images and deepfake videos, existing methods can be categorized based on their approach to extracting features, which include spatial pattern based~\cite{wang2020cnn,schwarcz2021finding, ju2022fusing, wang2023dire, ojha2023towards}, 
frequency analysis~\cite{gragnaniello2021gan,frank2020leveraging,corvi2023detection,corvi2023intriguing}, 
and spatial-temporal inconsistencies~\cite{choi2024exploiting,haliassos2022leveraging,gu2022delving,hu2022finfer}. 
In fake audio detection, extensive research has focused on speech synthesis and replay attack detection~\cite{wu2015spoofing, patil2018survey}. A variety of discriminative representations have been explored, including bi-spectral patterns~\cite{albadawy2019detecting}, Linear Frequency Cepstral Coefficients (LFCC) ~\cite{todisco2019asvspoof}, RawNet2 features~\cite{rawnet2}, vocoder artifacts~\cite{sun2023ai}, and Whisper features~\cite{kawa23b_interspeech}.


Beyond open-source detection methods, several online platforms and plugin modules have been developed to provide user-friendly tools to analyze digital media. These services make DeepFake detection accessible to everyone without requiring coding or research expertise.
Most of them are commercial tools with self-designed and closed-source detectors, such as 
{\it Deepware}~\cite{Deepware}, {\it WeVerify}~\cite{weverify}, {\it AI Voice Detector}~\cite{aivoice}, {\it DuckDuckGoose}~\cite{duckduckgoose}, {\it Sensity}~\cite{sensity}, {\it Resemble.AI}~\cite{resembleai}. 
Differently, our DeepFake-O-Meter v2.0 platform provides non-profit services and integrates multiple state-of-the-art detectors covering DeepFake image, video, and audio detection.
A detailed comparison of our platform with existing tools is summarized in Table~\ref{table:platforms}.

Our work expands upon the initial version of DeepFake-O-Meter v1.0 presented in~\cite{li2021deepfake}. This new version introduces the following three key improvements. First, instead of only targeting deepfake video detection, the DeepFake-O-Meter v2.0 adds multiple algorithms for DeepFake image and audio detection. Second, we re-design both the front and back end to provide more user-friendly and computation-efficient functions. Third, we emphasize the scalability of the platform by adding user feedback, usage analysis, and third-party docker creation and submission.

\begin{table*}[!h]
\center
\caption{Summary of existing platforms for DeepFake detection. ``-'' indicates no detectors, and ``N/A" indicates unknown information.}
\vspace{-0.12cm}
\label{table:platforms}
\scalebox{0.8}{
\begin{tabular}{|c|cc|ccc|c|}
\toprule[1.2pt]
\multirow{2}{*}{\textbf{Platform}} & \multicolumn{2}{c|}{\textbf{Usage}}         & \multicolumn{3}{c|}{\textbf{\# Detector}}       & \multirow{2}{*}{\textbf{Website}} \\ \cline{2-6} & \textbf{Open?} & \textbf{Free?} & \textbf{Image} & \textbf{Video} & \textbf{Audio}  & \\ 
\midrule[1.2pt]
Deepware 
& \ding{51} & \ding{51} & -&4&- &\url{https://scanner.deepware.ai/}\\ \hline
WeVerify
& \ding{51}&\ding{51} & N/A&N/A&- 
&\url{https://weverify.eu/verification-plugin/} \\ \hline
AI or Not & \ding{55} & \ding{55} & N/A & - & N/A & \url{https://www.aiornot.com/} \\ \hline
Content at Scale& \ding{55} & \ding{51} & N/A & - & - & \url{https://contentatscale.ai/ai-image-detector/} \\ \hline
AI Voice Detector
& \ding{55} &\ding{55} & -&-&N/A & \url{https://aivoicedetector.com/} \\ \hline
DuckDuckGoose 
& \ding{55}&\ding{55} & N/A&N/A&N/A & \url{https://duckduckgoose.ai/} \\ \hline
Sensity
& \ding{55}&\ding{55} & N/A&N/A&N/A & \url{https://sensity.ai/deepfake-detection/} \\ \hline
Resemble.AI
& \ding{55}&\ding{51} & -&-&N/A & \url{https://resemble.ai/free-deepfake-detector/} \\ \hline
TrueMedia
& \ding{51}&\ding{51}$^{\ddagger}$ & 8 & 5 & 3 & \url{https://truemedia.org/} \\ \hline
FakeCatcher (Intel)
& \ding{55}&\ding{55} & -&N/A&- &   \href{https://www.intel.com/content/www/us/en/research/trusted-media-deepfake-detection.html}{\nolinkurl{FakeCatcher}}\\ \hline
Video Authenticator (Microsoft)
& \ding{55}&\ding{55} & N/A&N/A&-  & 
\href{https://blogs.microsoft.com/on-the-issues/2020/09/01/disinformation-deepfakes-newsguard-video-authenticator/}{\nolinkurl{Video Authenticator}} \\
\hline
Pindrop Pulse
& \ding{55}&\ding{55} & -&-&N/A  & \url{https://www.pindrop.com/deepfake/} \\ \hline
Sentinel
& \ding{55}&\ding{55} & N/A&N/A&N/A  & \url{https://thesentinel.ai/} \\ \hline
DeepFake-o-meter v1.0 (Ours) & \ding{51}&\ding{51} & 2 & 9 & -  & -\\ \midrule[1.2pt]
DeepFake-o-meter v2.0 (Ours) & \ding{51}&\ding{51} & 6&6&5 & \url{http://zinc.cse.buffalo.edu/ubmdfl/deep-o-meter/home_login} \\ 
\hline
\end{tabular}
\vspace{-0.7cm}
}
\end{table*}


Our main contributions can be summarized as follows:
\begin{itemize}
    \item We design a new front-end architecture with user-friendly functions, including file upload supporting three media modalities, detector selection with 18 detectors, user information input, and authorized data collection.
    \item In the back-end design, we expand the detectors in DeepFake-O-Meter v1.0~\cite{li2021deepfake} to cover AI-generated image and audio detection with more advanced detection models. We also incorporate a job balancing module to help address conflicts between multiple users and tasks. 
    \item We conduct comprehensive statistical analysis to gain deep insights into the usage of our platform, including four-month user activity trends, detector processing efficiency, and query popularity. 
\end{itemize}

\vspace{-0.2cm}
\section{Platform Design}
\vspace{-0.1cm}
In this section, we describe the architecture design of the DeepFake-O-Meter v2.0 platform, which includes the front-end development of the user interface and back-end design with server-side components for data storage, detection processing, and application functionality.

\subsection {Front-end Design}
To interact with users, we built a website as the front end of the DeepFake-O-Meter v2.0 platform, allowing users to submit detection tasks, check the results, and manage user data. 
Our front end is built on a variety of technologies. The core of our system uses the Python Flask framework for logic processing and the Apache server for hosting and managing web services. The design and layout of the web pages are crafted using HTML and CSS. At the same time, AJAX technology enables asynchronous page updates, allowing for real-time updates for detection results without the need to reload the entire page. User information is stored in a SQLite database and includes encryption algorithms.

In this section, we will describe the whole process and provide details on the implementation of user interaction. 

\begingroup
\renewcommand\thefootnote{$\ddagger$}\footnotetext{Access requires a user application.}
\endgroup

\subsubsection{Account System}
We have developed an account system to safeguard user data and ensure secure usage of our platform. Users must log in before they can submit tasks. First-time users can see the login button on the homepage, click to enter the login page, and click Signup to register an account. Users need to fill in their username, email address, and password and verify the authenticity of their email address with the verification code sent by our system to complete the registration and log in for the first time. 


\begin{figure*}
    \centering
    \includegraphics[width=0.9\textwidth]{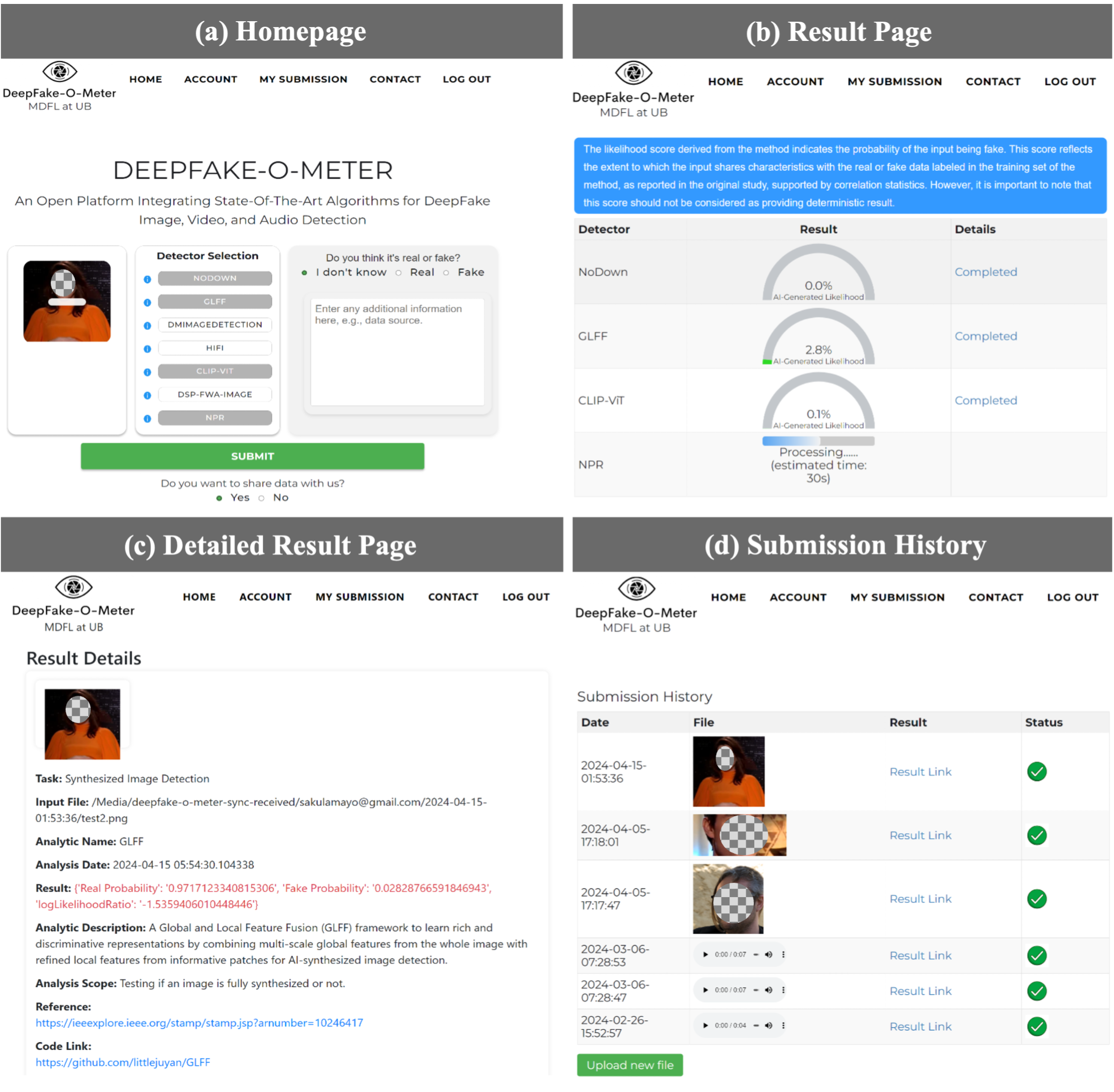}
    \vspace{-0.15cm}
    \caption{{DeepFake-O-Meter web pages, (a) Homepage, (b) Result Page, (c) Detailed Result Page, (d) Submission History Page. }}
    \vspace{-0.5cm}
    \label{fig:frontpages}
\end{figure*}

\subsubsection{Task Submission}
The homepage serves as the sole interface for users to submit tasks, which comprises three elements: a media file, detector selection, and supplementary information, as shown in Fig.~\ref{fig:frontpages} (a).

\noindent\textbf{File Upload.} To select a file to upload, users can either click on the file upload area and select a local file or drag and drop a file into this area. Users can change the file anytime by clicking or dragging it in again before submission. The format of the video, image, and audio that can be uploaded are shown in Table.~\ref{tab:filetype}. 
Moreover, we have set up a tiered user system for efficient resource allocation and enhanced server security, which includes regular, advanced, and super users, each with specific daily task submission limits. Super users, usually administrators, can submit tasks without any restrictions. Advanced users can submit up to 300 daily tasks, while regular users are limited to 30. 

\begin{table}[t]
\small
        \caption{Supported file type for task submission.}
\vspace{-0.12cm}
        \label{tab:filetype}
        \centering
        \begin{tabular}{cc}           
\toprule[1.2pt] \textbf{Media Modality} & \textbf{Supported File Type}\\ 
\midrule[1.2pt]
             Video& mp4, bmp, tif, nef, raf, avi, mov\\ \hline
             Image& jpg, png, jpeg\\ \hline
             Audio& flac, wav, mp3\\ 
\bottomrule[1.2pt]
        \end{tabular}
\vspace{-0.7cm}
    \end{table}
    
\noindent\textbf{Detector Selection.} Once a file is uploaded, the system will automatically present the available detection methods in the ``Detector Selection’’ module based on the file type. Users can choose one or multiple submission methods. Additionally, they can click the icon next to the method name to access the detector references and publicly available codes for a deep understanding.

    
\noindent\textbf{Supplementary Information.} 
Before finalizing the submission, we have designed the supplementary information module for users to provide additional information on the sample, such as data source, ground truth, and background information. Additionally, users can choose whether they want to share their data for academic research purposes only. This module is designed to support the ongoing advancement of the DeepFake-O-Meter v2.0 platform. Gathering user-granted data and background information about uploaded samples is crucial for this purpose. It serves as an effective method for collecting diverse DeepFake data, which can be utilized to enhance the detection methods on our platform in the future.

    

\subsubsection{Result Display}
After submitting the task, users are directed to the result page. Each selected detector displays a progress bar with an estimated average processing time while running on the back end. The front end actively tracks the processing progress in real time and promptly shows the results on the webpage. Each detector will output a likelihood probability of the uploaded sample, as depicted in Fig.~\ref{fig:frontpages} (b). The likelihood score derived from the method indicates the probability of the input being fake. This score reflects the extent to which the input shares characteristics with the real or fake data labeled in the training set of the method, as reported in the original study, supported by correlation statistics. However, it is important to note that this score should not be considered as providing a deterministic result. Additionally, we provide result links under the ``Completed'' tab for users to access a comprehensive detection report, which includes detailed prediction scores, method descriptions, paper references, source codes, and advanced results (if available, such as intermediate results and frame-level analysis for video detectors). One example is shown in Fig.~\ref{fig:frontpages} (c). 


\subsubsection{Submission History}

The submission history page, accessible through ``My Submission'' in the menu bar upon login, displays users' past submissions (refer to Fig.~\ref{fig:frontpages} (d)). It features a four-column table showcasing each task's submission time, sample preview, result link, and processing status.


\vspace{-0.15cm}
\subsection{Back-end Design}

Our back-end is a computation server with eight A5000 GPUs, which is mainly used for performing DeepFake detection methods. This section will provide an overview of the backend design, container building, and job balancing module. 

\subsubsection{Overview}
Once the user submits a file with selected detectors from the front end, the back end calls corresponding detection methods for the uploaded file. Since different detection methods rely on diverse environment settings and are tailored to different data formats, preprocessing steps, and result analyses, we have developed a unified framework to seamlessly integrate the state-of-the-art DeepFake detection algorithms. 
Specifically, our framework has two major designs: container creation and job balancing. 
The following steps introduce the logic of our back-end pipeline. 

\setlength\itemindent{0em}
\begin{enumerate}[leftmargin=*]
    \item[(1)] Check if any new task is submitted by monitoring the back-end task folder.
    \item[(2)] Parse the submitted task and extract the required information, such as username, file path, and submission time, for the next job balancing.
    \item[(3)] Sort jobs with a pre-defined priority, inversely proportional to the query frequency. This approach ensures fair distribution among users rather than favoring those with the highest query frequency. Jobs are then queued according to their priority. 
    \item[(4)] Constantly check if there are jobs in the queue and the GPU resources are available. If so, pop out the job and the GPU index.
    \item[(5)] Parse the corresponding detection method the user uploaded and run the corresponding docker container.
    \item[(6)] After detection, results will be shared back to the front end. 
\end{enumerate}

\subsubsection{Container Building} Virtual machines~\cite{li2021deepfake} were the initial solution to resolve environmental conflicts on a single machine. However, they are resource-intensive, involve redundant operations, and have slow startup times. Differently, containers~\cite{li2021deepfake} can isolate the process without simulating an entire operating system. Docker, the most popular container solution, allows developers to package applications and their dependencies into portable containers that can run on any machine. To facilitate the independent execution of each method, we create a separate Docker image for each detection method. Our platform integrates the following DeepFake detector, including 6 image detection methods, 7 video detection methods, and 5 audio detection methods. Each detection docker receives an image, video, or audio file as input and outputs detection probabilities and advanced results (if available). These results are then sent to the front-end for users to review. The summary of each detection method is given in Table~\ref{table:detectionmethod}.
\begin{table*}[!h]
\footnotesize
\center
\caption{Details of integrated open-source DeepFake detection methods.}
\vspace{-0.22cm}
\label{table:detectionmethod}
\newcommand{\tabincell}[2]{\begin{tabular}{@{}#1@{}}#2\end{tabular}} 
\scalebox{0.8}{
\begin{tabular}{|l|l|l|l|l|}
\toprule[1.2pt]
\textbf{Methods }                                                & \textbf{Year} & \textbf{Organization} & \textbf{Detection Scope}                                                       & \textbf{Repositories}                                                          \\ \midrule[1.0pt]
Nodown \cite{gragnaniello2021gan}                                               & 2021     & University Federico II of Naples   & \tabincell{l}{Image\\ (trained on StyleGAN2)}                    & \url{https://github.com/grip-unina/GANimageDetection}                         \\ \hline
GLFF \cite{ju2023glff}                                                   & 2023     &University at Buffalo        & \tabincell{l}{Image \\(trained on ProGAN and\\DALL-E images)} & \url{https://github.com/littlejuyan/FusingGlobalandLocal}                     \\ \hline
HIFI \cite{guo2023hierarchical}                                                    & 2023    &Michigan State University        & \tabincell{l}{Image \\(trained on GAN and diffusion)}            & \url{https://github.com/CHELSEA234/HiFi\_IFDL}                              \\ \hline
\tabincell{l}{DMimage-\\Detection} \cite{corvi2023detection}                                        & 2023    & University Federico II of Naples        & \tabincell{l}{Image\\ (trained on ProGAN and LDM) }              & \url{https://github.com/grip-unina/DMimageDetection}                          \\ \hline
CLIP-ViT \cite{ojha2023towards}                                                & 2023    &University of Wisconsin-Madison         & \tabincell{l}{Image\\ (trained on ProGAN)}                       & \url{https://github.com/Yuheng-Li/UniversalFakeDetect}                        \\  \hline
NPR \cite{tan2023rethinking}                                                & 2024    & Beijing Jiaotong University        & \tabincell{l}{Image \\(trained on ProGAN)}                       & \url{https://github.com/chuangchuangtan/NPR-DeepfakeDetection}                        \\ 
\midrule[1.2pt]
DSP-FWA \cite{li2019exposing}                                                  & 2019   & University at Albany         & \tabincell{l}{Face-swap deepfake \\ image and video}                                        & \url{https://github.com/yuezunli/DSP-FWA}                                     \\ \hline
FTCN \cite{zheng2021exploring}                                                 & 2021  & Xiamen University          & Face-swap deepfake video                                      & \url{https://github.com/yinglinzheng/FTCN}                                                   \\ \hline  
Wav2lip-STA \cite{jia2022model}                                             & 2022    & University at Buffalo        & Lip-syncing deepfake video                                      & \url{https://github.com/shanface33/Deepfake\_Model\_Attribution}              \\ \hline
SBI \cite{shiohara2022detecting}                            & 2022   & The University of Tokyo         & Face-swap deepfake video                                          & \url{https://github.com/mapooon/SelfBlendedImages}                            \\ \hline
AltFreezing \cite{Wang_2023_CVPR}                      & 2023     & \tabincell{l}{University of Science\\ and Technology of China}       & Face-swap deepfake video                                        & \url{https://github.com/ZhendongWang6/AltFreezing}                            \\ \hline
LIPINC \cite{datta2024exposing}                                                   & 2024     & University at Buffalo       & Lip-syncing deepfake video                                        & \url{https://github.com/skrantidatta/LIPINC} 
                           \\ \hline
LSDA \cite{yan2024transcending}                      & 2024     & \tabincell{l}{The Chinese University of \\ Hong Kong}       & Face-swap deepfake video                                        & \url{https://github.com/SCLBD/DeepfakeBench/tree/main}                         
                           \\  \midrule[1.2pt]
RawNet2 \cite{tak2021end}                                                    & 2021    &  Eurecom        & \tabincell{l}{Audio \\(trained on ASVspoof 2019 LA) }               & \url{https://github.com/eurecom-asp/rawnet2-antispoofing}                                                          \\ \hline
LFCC-LCNN \cite{wang2021comparative}  & 2021    &   National Institute of Informatics       & \tabincell{l}{Audio \\(trained on ASVspoof 2021 DF)}     & \url{https://github.com/nii-yamagishilab/project-NN-Pytorch-scripts/tree/master/project}                                                          \\ \hline
RawNet3 \cite{jung2022pushing}                                                  & 2022    &Naver Corporation           & \tabincell{l}{Audio \\(trained on ASVspoof 2021 DF)}                                     & \url{https://github.com/Jungjee/RawNet}                                                         \\ \hline
\tabincell{l}{RawNet2-\\Vocoder} \cite{sun2023ai}                                          & 2023    &University at Buffalo         & \tabincell{l}{Audio \\(trained on LibriSeVoc)}                   & \url{https://github.com/csun22/Synthetic-Voice-Detection-Vocoder-Artifacts}   \\ \hline
Whisper \cite{kawa23b_interspeech}                                                  & 2023    & \tabincell{l}{Wrocław University of \\ Science and Technology}        & \tabincell{l}{Audio \\(trained on ASVspoof 2021 DF)}                & \url{https://github.com/piotrkawa/deepfake-whisper-features}               \\ \bottomrule[1.2pt]

\end{tabular}
\vspace{-0.7cm}
}
\end{table*}
\setlength\itemindent{0em}
\begin{itemize}[leftmargin=*]
    \item  Nodown~\cite{gragnaniello2021gan}: This work introduces DeepFake image detectors without performing down-sampling in the first layer on ResNet50, XceptionNet, and Efficient-B4 backbones. 
    \item GLFF~\cite{ju2023glff}: This work introduces a two-branch model to improve the generalization ability of DeepFake image detection. It combines global spatial information from the whole image and local informative features from multiple patches selected by a novel patch selection module.

    \item HIFI~\cite{guo2023hierarchical}: The method contains three components: multi-branch feature extractor, localization, and classification modules. Each branch of the feature extractor learns to classify forgery attributes at one level, while localization and classification modules segment the pixel-level forgery region and detect image-level forgery, respectively. 
    
    \item DMImageDetection~\cite{corvi2023detection}: 
    This work averages the outputs of several GAN image detectors trained on GAN and Diffusion images. To improve the accuracy, a calibration procedure using just two real images and two synthetic ones for each model is used. 
    
    \item CLIP-ViT~\cite{ojha2023towards}: This work proposes to perform real-vs-fake image classification without learning, i.e., using a feature space not explicitly trained to distinguish real from fake images. Based on the feature space of a large pre-trained vision-language model (CLIP-ViT), nearest neighbor and linear probing are used for classification. It achieves good generalization ability in detecting fake images from various generative models.

    \item NPR~\cite{tan2023rethinking}: This work finds that the local interdependence among image pixels caused by up-sampling operators is significantly demonstrated in synthetic images generated by GAN or diffusion. Based on this, they present Neighboring Pixel Relationships (NPR) to achieve generalized deepfake detection.
    
    \item DSP-FWA~\cite{li2019exposing}: This work 
    is designed to detect face-swap DeepFake videos. It is based on the observations that commonly-used transform leave certain distinctive artifacts in the resulting DeepFake Videos, which a dedicated deep neural network model can capture.
        
    \item WAV2Lip-STA~\cite{jia2022model}: This work proposes a model attribution method by fusing spatial (frame-level) and temporal (video-level) attention schemes for discriminative feature extraction. ResNet-50 is used as the CNN feature extractor. It is fine-tuned on lip-syncing DeepFake videos.
    
    \item FTCN~\cite{zheng2021exploring}: 
    This method proposes a fully temporal convolution network (FTCN) to reduce the spatial convolution kernel size to one while maintaining the temporal convolution kernel size unchanged. It also designs a temporal transformer network to explore long-term temporal coherence for DeepFake video detection. 
    
    \item SBI~\cite{shiohara2022detecting}: This work presents novel self-blended images (SBIs) as training data to train face-swap DeepFake detectors. The idea is that more general and hardly recognizable fake samples encourage classifiers to learn generic and robust representations without overfitting to manipulation-specific artifacts. It adopts the EfficientNet-b4 network pre-trained on ImageNet as the classifier. 
    
    \item AltFreezing~\cite{Wang_2023_CVPR}: This work uses 3D ResNet50 as backbone for more general face forgery detection. It trains the model with the proposed AltFreezing strategy that separates the spatial and temporal weights into two groups and alternately freezes one group of weights to encourage the model to capture both the spatial and temporal artifacts.
    \item LSDA~\cite{yan2024transcending}: This work designs a simple yet effective detector called Latent Space Data Augmentation (LSDA), based on the idea that representations with a wider variety of forgeries should be able to learn a more generalizable decision boundary, thereby mitigating the overfitting of method-specific features. The method enlarges the forgery space by constructing and simulating variations within and across forgery features in the latent space for generalizable DeepFake video detection.
    \item LIPINC~\cite{datta2024exposing}: This detector targets lip-syncing DeepFake detection by identifying temporal inconsistencies in the mouth region. 
    It involves a local and global mouth frame extractor to extract adjacent and similarly posed mouth frames based on mouth openness throughout the video sequence and a spatial-temporal inconsistency extractor for encoding and learning distinctive inconsistency features. 
    
    \item RawNet2 ~\cite{tak2021end}: 
    This method applies the RawNet2 architecture to audio anti-spoofing tasks. It fuses the RawNet2 classifier and the high-spectral resolution Linear Frequency Cepstral Coefficient (LFCC) classifier to enhance the audio anti-spoofing performance.
    
    \item RawNet3~\cite{jung2022pushing}: This detector applies a novel speaker recognition model, RawNet3, based on raw waveform inputs for fake audio detection. It incorporates the Res2Net backbone and multi-layer feature aggregation.
    
    \item LFCC-LCNN~\cite{wang2021comparative}: 
    This method integrates LFCC feature extraction with an LCNN (Light Convolutional Neural Network) classifier for speech deepfake detection. 

    \item RawNet2-Vocoder~\cite{sun2023ai}: This detector proposes a multi-task learning framework for identifying synthetic human voices. It uses a binary-class RawNet2 model that shares the feature extractor with a vocoder identification module. By treating vocoder identification as a pretext task, the method constrains the feature extractor from focusing on vocoder artifacts for deepfake voice detection.
    
    \item Whisper~\cite{kawa23b_interspeech}: This work uses the Whisper encoder as a feature extractor in deepfake audio detection and performs well when combined with the existing detection architectures. 
    
\end{itemize}


\subsubsection{Job Balancing} 
\begin{figure}[t]
\centering
\includegraphics[scale=0.24]{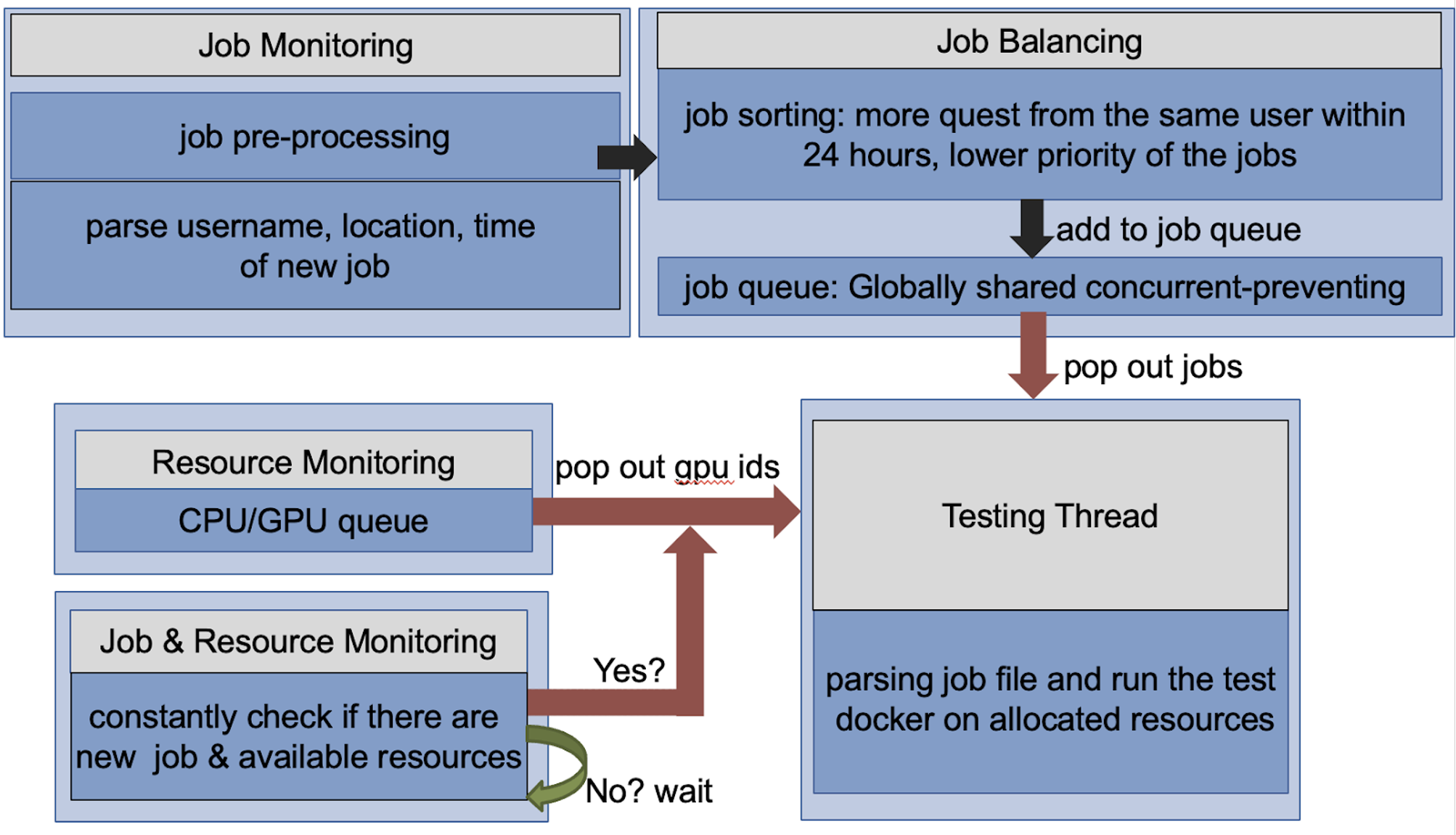}
\vspace{-0.12cm}
\caption{Pipeline of the job balancing module.}
\label{fig:backend1}
\vspace{-0.52cm}
\end{figure}

{We design a monitoring thread that parses username, file location, uploading time information when a new task is submitted. Details of the job balancing module is described in Fig.~\ref{fig:backend1}.}

\begin{figure*}[th]
    \centering
    \includegraphics[width=0.9\linewidth]{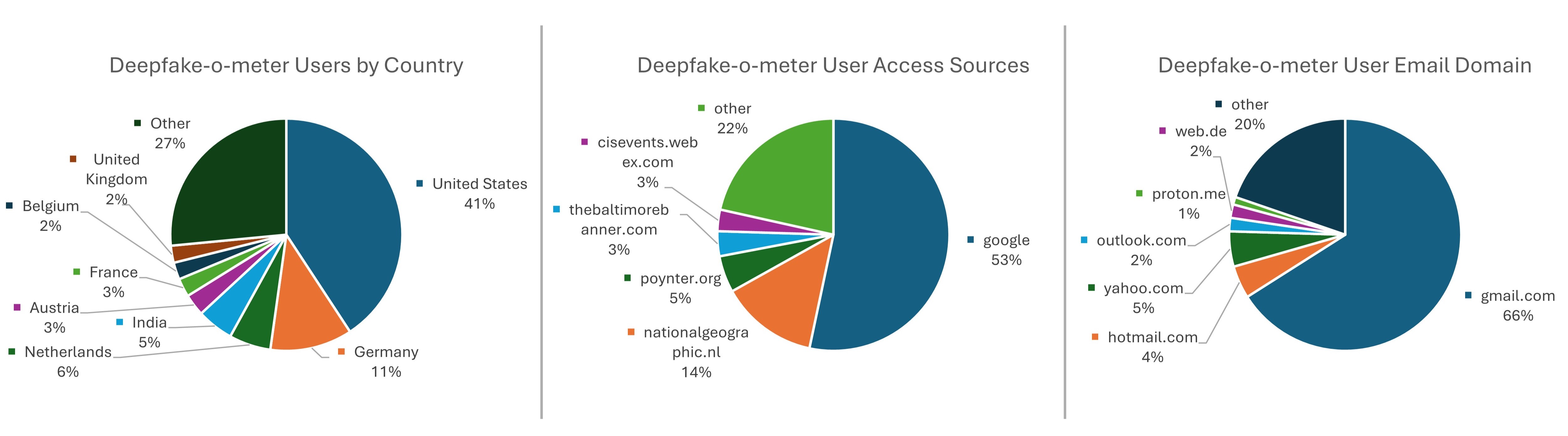}
\vspace{-0.32cm}
    \caption{User pattern statistics based on data from February 8, 2024, to June 1, 2024.}
    \label{fig:user_statistics}
\vspace{-0.62cm}
\end{figure*}

\section{Platform Data Analysis}
In this section, we dive into usage statistics to reveal insights about our platform. Based on our dataset spanning 115 days of user activity, we first analyze usage trends and then conduct an in-depth detailed analysis of each detector's processing efficiency and query popularity.

\subsection{Usage Analysis}

\begin{figure}[tbp]
    \begin{minipage}[t]{1.0\linewidth}
        \centering
        \includegraphics[width=1\textwidth]{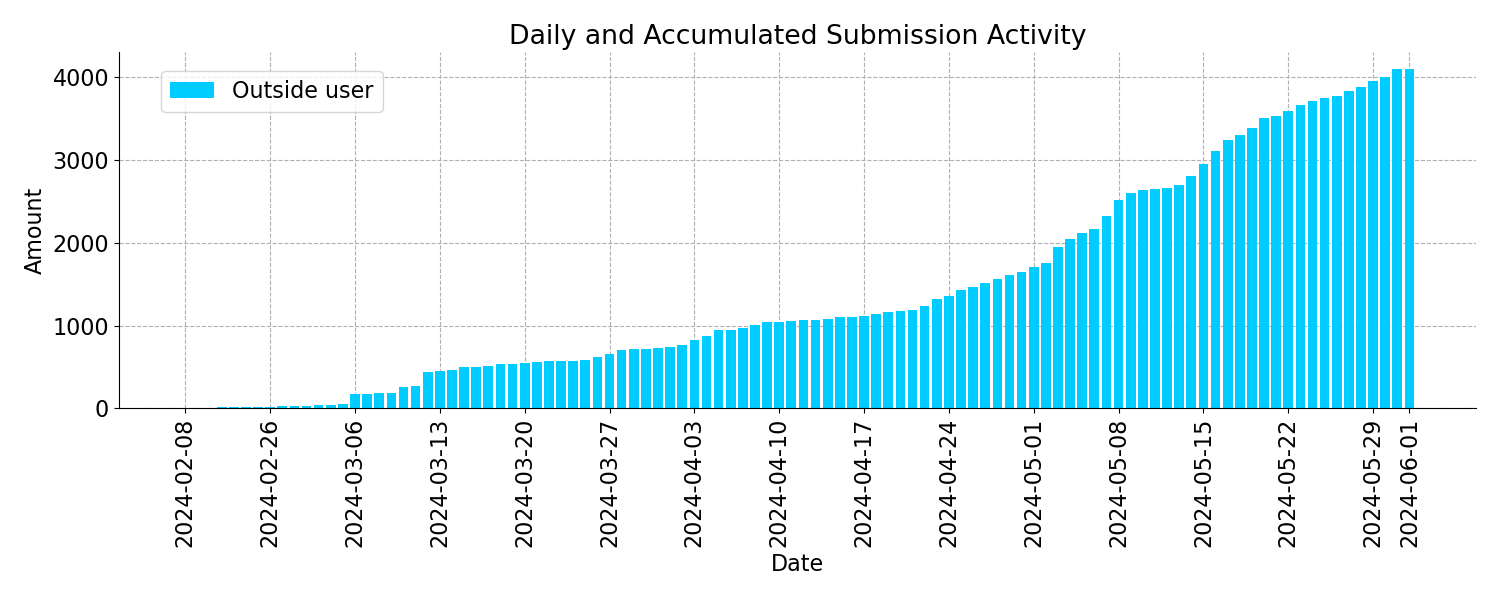}
    \end{minipage}
    \vspace{-0.52cm}
    \caption{Accumulative daily submission activity of our platform.}
    \vspace{-0.52cm}
    \label{fig:daily}
\end{figure}

In this section, we show the platform's usage activities collected from February 8, 2024, to June 1, 2024. The platform currently has 632 registered users, among which 445 users submitted 4,091 tasks during this period. In Fig.~\ref{fig:user_statistics}, we provide statistical analysis of the user demographics, including the countries and regions, as well as the browsing behavior, such as access sources and patterns in their registered email domains. The accesses to the platform come from 98 countries, with the largest portion from the United States (41\%) and Germany (11\%). Regarding access sources, 53\% of users reached our platform via Google searches, while the remaining half came through referrals from various websites, including news pages like National Geographic. This suggests that the proliferation and media coverage of DeepFake technology have notably impacted our website traffic. Furthermore, the email addresses of the platform users spread across 296 domains, among which 66\% of users registered on our platform using Gmail accounts. Fig. \ref{fig:daily} shows the accumulative submission activity of the platform during the data collection period. The figure shows that our platform has experienced significant submission growth since May 1, 2024, and there are, on average, 34 new submissions each day.

\vspace{-0.2cm}
\subsection{Detector Analysis}



\begin{figure}[t]
\centering
\includegraphics[scale=0.18]{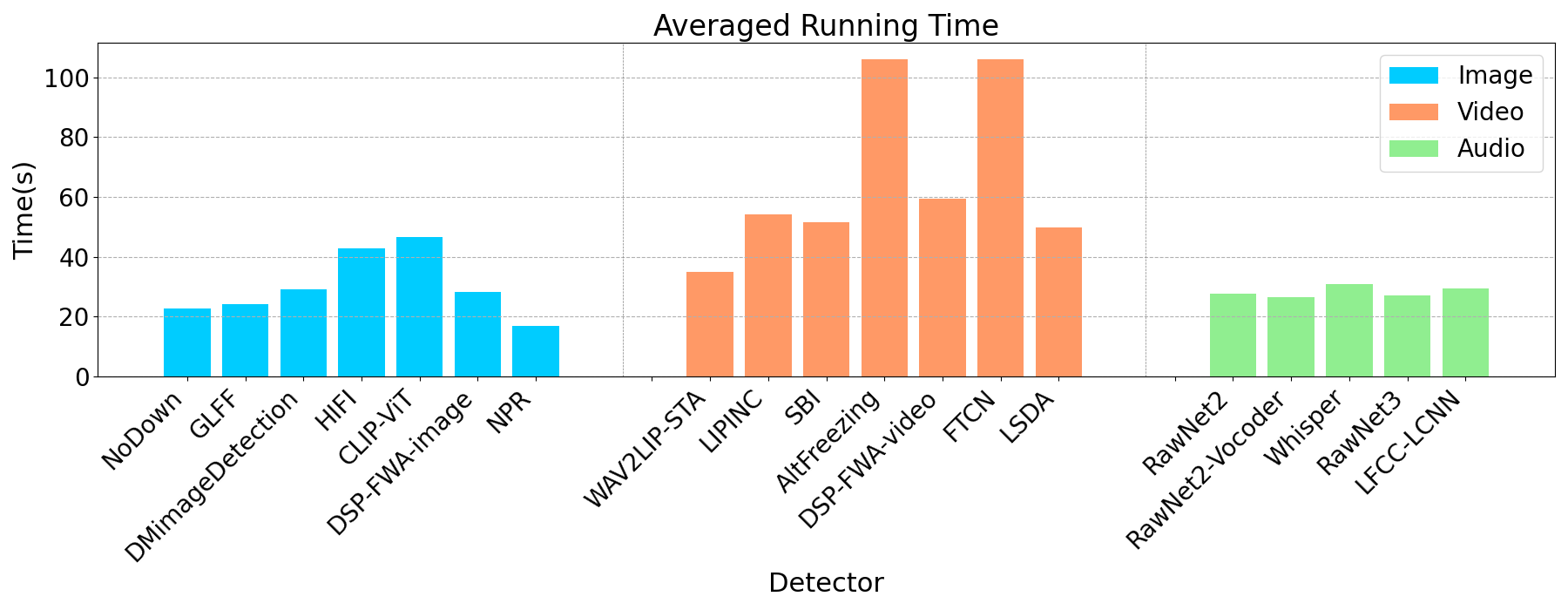}
\vspace{-0.12cm}
\caption{Running time of each module}
\label{fig:runningtime}
\vspace{-0.32cm}
\end{figure}

\begin{figure}[t]
\centering
\includegraphics[scale=0.18]{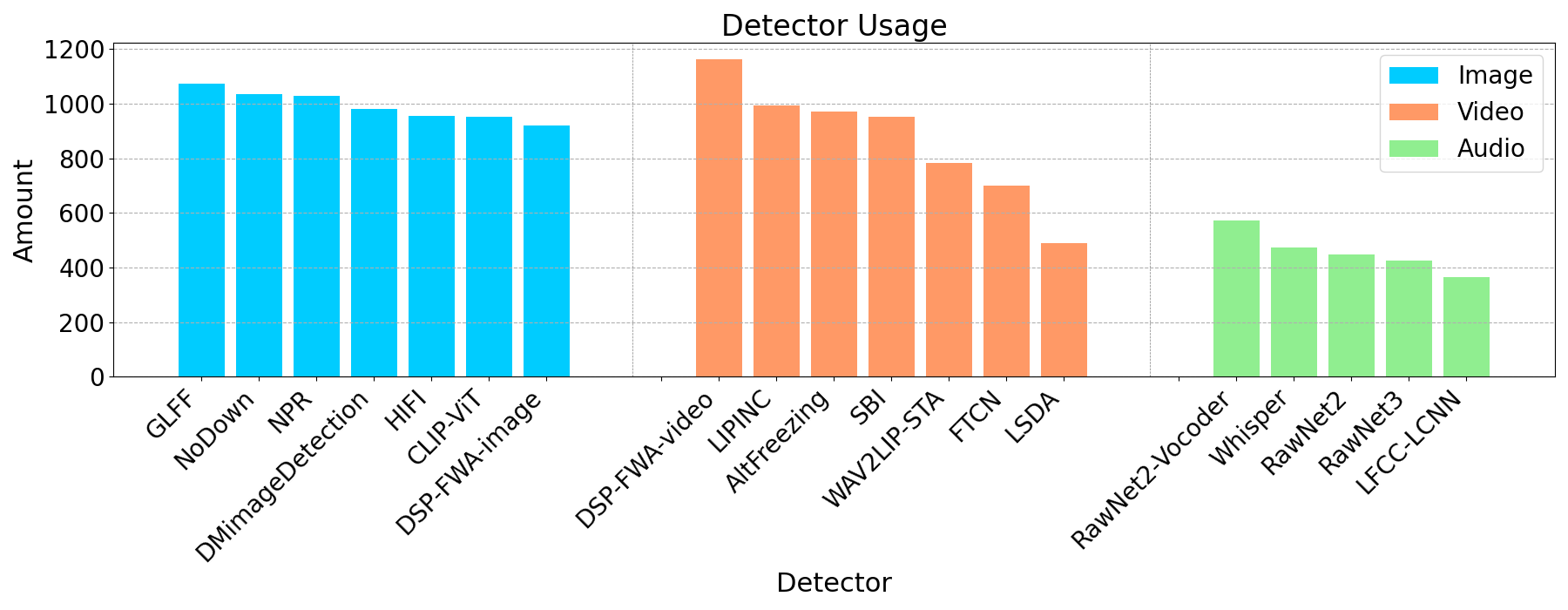}
\vspace{-0.12cm}
\caption{Number of query of each detector.}
\label{fig:querynumber}
\vspace{-0.52cm}
\end{figure}

To evaluate the efficiency of the back-end testing, we recorded the running time for each detection task across various modules and presented the average running time for each detection module in Fig.~\ref{fig:runningtime}. The figure reveals that the average running time for image and audio detection modules is approximately 30 seconds. For video modules, the average time extends to approximately 90 seconds, primarily due to the intricate processes involved in face cropping and frame-by-frame prediction. Furthermore, we assess the usage popularity of each detector in Fig.~\ref{fig:querynumber}. 
The figure shows that the use of image and video detectors is significantly more widespread than that of audio detectors, indicating that the appearance of potential fake images and fake videos are more prevailing.

\vspace{-0.3cm}
\section{Conclusion}
In this paper, we introduce the DeepFake-O-Meter v2.0, an open platform that integrates multiple state-of-the-art algorithms for detecting DeepFake images, videos, and audio. Our platform aims to offer comprehensive, user-friendly, and accessible services to a wide audience, including academic researchers and the general public.

There are several directions for us to improve our platform in the future. Firstly, we will focus on continually integrating advanced detectors and enhancing the interface for developers to submit detection algorithms conveniently. We also plan to explore the inclusion of multi-modal detection methods, such as audio-visual DeepFake video detectors~\cite{muppalla2023integrating, cozzolino2023audio} and text-image inconsistency detectors \cite{abdelnabi2022open, huang2023exposing}. Furthermore, enhancing the detection efficiency is also a crucial direction for improvement. We will also conduct thorough performance evaluations of integrated detectors using real-world data and offer detailed insights to assist users in selecting the most suitable detectors for their needs on our platform. 

\noindent\textbf{Acknowledgement.} This work was supported in part by the US Defense Advanced Research Projects Agency (DARPA) Semantic Forensic (SemaFor) program, under Contract No. HR001120C0123, National Science Foundation (NSF) Projects under grants SaTC-2153112, No.1822190, and TIP-2137871, and University at Buffalo's Office of Vice President for Research and Economic Development and Center for Information Integrity. The views and conclusions contained herein are those of the authors and should not be interpreted as necessarily representing the official policies, either expressed or implied, of DARPA, NSF, or the U.S. Government.

%
\IEEEpeerreviewmaketitle

\bibliographystyle{IEEEtran}
\footnotesize
\bibliography{main}

\begin{thebibliography}{10}
\providecommand{\url}[1]{#1}
\csname url@samestyle\endcsname
\providecommand{\newblock}{\relax}
\providecommand{\bibinfo}[2]{#2}
\providecommand{\BIBentrySTDinterwordspacing}{\spaceskip=0pt\relax}
\providecommand{\BIBentryALTinterwordstretchfactor}{4}
\providecommand{\BIBentryALTinterwordspacing}{\spaceskip=\fontdimen2\font plus
\BIBentryALTinterwordstretchfactor\fontdimen3\font minus \fontdimen4\font\relax}
\providecommand{\BIBforeignlanguage}[2]{{%
\expandafter\ifx\csname l@#1\endcsname\relax
\typeout{** WARNING: IEEEtran.bst: No hyphenation pattern has been}%
\typeout{** loaded for the language `#1'. Using the pattern for}%
\typeout{** the default language instead.}%
\else
\language=\csname l@#1\endcsname
\fi
#2}}
\providecommand{\BIBdecl}{\relax}
\BIBdecl

\bibitem{karras2019style}
T.~Karras, S.~Laine, and T.~Aila, ``A style-based generator architecture for generative adversarial networks,'' in \emph{CVPR}, 2019, pp. 4401--4410.

\bibitem{ho2020denoising}
J.~Ho, A.~Jain, and P.~Abbeel, ``Denoising diffusion probabilistic models,'' \emph{Advances in Neural Information Processing Systems}, vol.~33, pp. 6840--6851, 2020.

\bibitem{rombach2022high}
R.~Rombach, A.~Blattmann, D.~Lorenz, P.~Esser, and B.~Ommer, ``High-resolution image synthesis with latent diffusion models,'' in \emph{CVPR}, 2022, pp. 10\,684--10\,695.

\bibitem{MakePixelsDance}
Y.~Zeng, G.~Wei \emph{et~al.}, ``Make pixels dance: High-dynamic video generation,'' \emph{arXiv:2311.10982}, 2023.

\bibitem{cho2024sora}
J.~Cho, F.~D. Puspitasari, S.~Zheng, J.~Zheng, L.-H. Lee, T.-H. Kim, C.~S. Hong, and C.~Zhang, ``Sora as an agi world model? a complete survey on text-to-video generation,'' \emph{arXiv preprint arXiv:2403.05131}, 2024.

\bibitem{liu2023evalcrafter}
Y.~Liu, X.~Cun \emph{et~al.}, ``Evalcrafter: Benchmarking and evaluating large video generation models,'' \emph{arXiv:2310.11440}, 2023.

\bibitem{huang2023make}
R.~Huang, J.~Huang \emph{et~al.}, ``Make-an-audio: Text-to-audio generation with prompt-enhanced diffusion models,'' in \emph{International Conference on Machine Learning}.\hskip 1em plus 0.5em minus 0.4em\relax PMLR, 2023, pp. 13\,916--13\,932.

\bibitem{liu2023audioldm}
H.~Liu, Z.~Chen \emph{et~al.}, ``Audioldm: Text-to-audio generation with latent diffusion models,'' \emph{arXiv:2301.12503}, 2023.

\bibitem{kreuk2022audiogen}
F.~Kreuk, G.~Synnaeve \emph{et~al.}, ``Audiogen: Textually guided audio generation,'' \emph{arXiv preprint arXiv:2209.15352}, 2022.

\bibitem{borsos2023audiolm}
Z.~Borsos, R.~Marinier, D.~Vincent \emph{et~al.}, ``Audiolm: a language modeling approach to audio generation,'' \emph{IEEE/ACM Transactions on Audio, Speech, and Language Processing}, 2023.

\bibitem{lyu2020deepfake}
S.~Lyu, ``Deepfake detection: Current challenges and next steps,'' in \emph{2020 IEEE international conference on multimedia \& expo workshops (ICMEW)}.\hskip 1em plus 0.5em minus 0.4em\relax IEEE, 2020, pp. 1--6.

\bibitem{masood2023deepfakes}
M.~Masood, M.~Nawaz, K.~M. Malik, A.~Javed, A.~Irtaza, and H.~Malik, ``Deepfakes generation and detection: State-of-the-art, open challenges, countermeasures, and way forward,'' \emph{Applied intelligence}, vol.~53, no.~4, pp. 3974--4026, 2023.

\bibitem{khanjani2023audio}
Z.~Khanjani, G.~Watson, and V.~P. Janeja, ``Audio deepfakes: A survey,'' \emph{Frontiers in Big Data}, vol.~5, p. 1001063, 2023.

\bibitem{malik2022deepfake}
A.~Malik, M.~Kuribayashi, S.~M. Abdullahi, and A.~N. Khan, ``Deepfake detection for human face images and videos: A survey,'' \emph{Ieee Access}, vol.~10, pp. 18\,757--18\,775, 2022.

\bibitem{wang2020cnn}
S.-Y. Wang, O.~Wang \emph{et~al.}, ``Cnn-generated images are surprisingly easy to spot... for now,'' in \emph{CVPR}, 2020, pp. 8695--8704.

\bibitem{schwarcz2021finding}
S.~Schwarcz and R.~Chellappa, ``Finding facial forgery artifacts with parts-based detectors,'' in \emph{CVPR}, 2021, pp. 933--942.

\bibitem{ju2022fusing}
Y.~Ju, S.~Jia, L.~Ke, H.~Xue, K.~Nagano, and S.~Lyu, ``Fusing global and local features for generalized ai-synthesized image detection,'' in \emph{ICIP}.\hskip 1em plus 0.5em minus 0.4em\relax IEEE, 2022, pp. 3465--3469.

\bibitem{wang2023dire}
Z.~Wang, J.~Bao \emph{et~al.}, ``Dire for diffusion-generated image detection,'' in \emph{ICCV}, 2023, pp. 22\,445--22\,455.

\bibitem{ojha2023towards}
U.~Ojha, Y.~Li, and Y.~J. Lee, ``Towards universal fake image detectors that generalize across generative models,'' in \emph{CVPR}, 2023, pp. 24\,480--24\,489.

\bibitem{gragnaniello2021gan}
D.~Gragnaniello, D.~Cozzolino, F.~Marra, G.~Poggi, and L.~Verdoliva, ``Are gan generated images easy to detect? a critical analysis of the state-of-the-art,'' in \emph{ICME}.\hskip 1em plus 0.5em minus 0.4em\relax IEEE, 2021, pp. 1--6.

\bibitem{frank2020leveraging}
J.~Frank, T.~Eisenhofer \emph{et~al.}, ``Leveraging frequency analysis for deep fake image recognition,'' in \emph{ICML}.\hskip 1em plus 0.5em minus 0.4em\relax PMLR, 2020, pp. 3247--3258.

\bibitem{corvi2023detection}
R.~Corvi, D.~Cozzolino \emph{et~al.}, ``On the detection of synthetic images generated by diffusion models,'' in \emph{ICASSP}.\hskip 1em plus 0.5em minus 0.4em\relax IEEE, 2023, pp. 1--5.

\bibitem{corvi2023intriguing}
R.~Corvi, D.~Cozzolino, G.~Poggi, K.~Nagano, and L.~Verdoliva, ``Intriguing properties of synthetic images: from generative adversarial networks to diffusion models,'' in \emph{CVPR}, 2023, pp. 973--982.

\bibitem{choi2024exploiting}
J.~Choi, T.~Kim, Y.~Jeong, S.~Baek, and J.~Choi, ``Exploiting style latent flows for generalizing deepfake detection video detection,'' \emph{arXiv preprint arXiv:2403.06592}, 2024.

\bibitem{haliassos2022leveraging}
A.~Haliassos, R.~Mira, S.~Petridis, and M.~Pantic, ``Leveraging real talking faces via self-supervision for robust forgery detection,'' in \emph{CVPR}, 2022, pp. 14\,950--14\,962.

\bibitem{gu2022delving}
Z.~Gu, Y.~Chen, T.~Yao, S.~Ding, J.~Li, and L.~Ma, ``Delving into the local: Dynamic inconsistency learning for deepfake video detection,'' in \emph{AAAI}, vol.~36, no.~1, 2022, pp. 744--752.

\bibitem{hu2022finfer}
J.~Hu, X.~Liao, J.~Liang, W.~Zhou, and Z.~Qin, ``Finfer: Frame inference-based deepfake detection for high-visual-quality videos,'' in \emph{AAAI}, vol.~36, no.~1, 2022, pp. 951--959.

\bibitem{wu2015spoofing}
Z.~Wu, N.~Evans, T.~Kinnunen, J.~Yamagishi, F.~Alegre, and H.~Li, ``Spoofing and countermeasures for speaker verification: A survey,'' \emph{speech communication}, vol.~66, pp. 130--153, 2015.

\bibitem{patil2018survey}
H.~A. Patil and M.~R. Kamble, ``A survey on replay attack detection for automatic speaker verification (asv) system,'' in \emph{2018 Asia-Pacific Signal and Information Processing Association Annual Summit and Conference (APSIPA ASC)}.\hskip 1em plus 0.5em minus 0.4em\relax IEEE, 2018, pp. 1047--1053.

\bibitem{albadawy2019detecting}
E.~A. AlBadawy, S.~Lyu, and H.~Farid, ``Detecting {AI}-synthesized speech using bispectral analysis.'' in \emph{CVPR Workshops}, 2019, pp. 104--109.

\bibitem{todisco2019asvspoof}
M.~Todisco, X.~Wang \emph{et~al.}, ``{ASV}spoof 2019: Future horizons in spoofed and fake audio detection,'' \emph{arXiv preprint arXiv:1904.05441}, 2019.

\bibitem{rawnet2}
H.~Tak, J.~Patino \emph{et~al.}, ``End-to-end anti-spoofing with rawnet2,'' in \emph{ICASSP}, 2021, pp. 6369--6373.

\bibitem{sun2023ai}
C.~Sun, S.~Jia, S.~Hou, and S.~Lyu, ``Ai-synthesized voice detection using neural vocoder artifacts,'' in \emph{CVPR}, 2023, pp. 904--912.

\bibitem{kawa23b_interspeech}
P.~Kawa, M.~Plata \emph{et~al.}, ``{Improved DeepFake Detection Using Whisper Features},'' in \emph{Proc. INTERSPEECH 2023}, 2023, pp. 4009--4013.

\bibitem{Deepware}
Deepware. (2021) {Deepware}. Available at \url{https://scanner.deepware.ai/}.

\bibitem{weverify}
WeVerify. (2020) {WeVerify}. Available at \url{https://weverify.eu/verification-plugin//}.

\bibitem{aivoice}
AIVoiceDetector. (2020) {AI Voice Detector}. Available at \url{https://aivoicedetector.com//}.

\bibitem{duckduckgoose}
DuckDuckGoose. (2023) {DuckDuckGoose}. Available at \url{https://www.duckduckgoose.ai/}.

\bibitem{sensity}
Sensity.AI. (2023) {Sensity}. Available at \url{https://sensity.ai/deepfake-detection//}.

\bibitem{resembleai}
Resemble.AI. (2023) {Resemble AI}. Available at \url{https://www.resemble.ai/free-deepfake-detector//}.

\bibitem{li2021deepfake}
Y.~Li, C.~Zhang, P.~Sun, L.~Ke, Y.~Ju, H.~Qi, and S.~Lyu, ``Deepfake-o-meter: An open platform for deepfake detection,'' in \emph{2021 IEEE Security and Privacy Workshops (SPW)}.\hskip 1em plus 0.5em minus 0.4em\relax IEEE, 2021, pp. 277--281.

\bibitem{ju2023glff}
Y.~Ju, S.~Jia, J.~Cai, H.~Guan, and S.~Lyu, ``Glff: Global and local feature fusion for ai-synthesized image detection,'' \emph{IEEE Transactions on Multimedia}, 2023.

\bibitem{guo2023hierarchical}
X.~Guo, X.~Liu \emph{et~al.}, ``Hierarchical fine-grained image forgery detection and localization,'' in \emph{CVPR}, 2023, pp. 3155--3165.

\bibitem{tan2023rethinking}
C.~Tan, Y.~Zhao, S.~Wei, G.~Gu, P.~Liu, and Y.~Wei, ``Rethinking the up-sampling operations in cnn-based generative network for generalizable deepfake detection,'' \emph{arXiv preprint arXiv:2312.10461}, 2023.

\bibitem{li2019exposing}
Y.~Li and S.~Lyu, ``Exposing deepfake videos by detecting face warping artifacts,'' in \emph{CVPRW}, 2019.

\bibitem{zheng2021exploring}
Y.~Zheng, J.~Bao \emph{et~al.}, ``Exploring temporal coherence for more general video face forgery detection,'' in \emph{ICCV}, 2021, pp. 15\,044--15\,054.

\bibitem{jia2022model}
S.~Jia, X.~Li, and S.~Lyu, ``Model attribution of face-swap deepfake videos,'' in \emph{ICIP}.\hskip 1em plus 0.5em minus 0.4em\relax IEEE, 2022, pp. 2356--2360.

\bibitem{shiohara2022detecting}
K.~Shiohara and T.~Yamasaki, ``Detecting deepfakes with self-blended images,'' in \emph{CVPR}, 2022, pp. 18\,720--18\,729.

\bibitem{Wang_2023_CVPR}
Z.~Wang, J.~Bao, W.~Zhou, W.~Wang, and H.~Li, ``Altfreezing for more general video face forgery detection,'' in \emph{CVPR}, June 2023, pp. 4129--4138.

\bibitem{datta2024exposing}
S.~K. Datta, S.~Jia, and S.~Lyu, ``Exposing lip-syncing deepfakes from mouth inconsistencies,'' \emph{ICME}, 2024.

\bibitem{yan2024transcending}
Z.~Yan, Y.~Luo, S.~Lyu, Q.~Liu, and B.~Wu, ``Transcending forgery specificity with latent space augmentation for generalizable deepfake detection,'' in \emph{Proceedings of the IEEE/CVF Conference on Computer Vision and Pattern Recognition}, 2024, pp. 8984--8994.

\bibitem{tak2021end}
H.~Tak, J.~Patino, M.~Todisco, A.~Nautsch, N.~Evans, and A.~Larcher, ``End-to-end anti-spoofing with rawnet2,'' in \emph{ICASSP}.\hskip 1em plus 0.5em minus 0.4em\relax IEEE, 2021, pp. 6369--6373.

\bibitem{wang2021comparative}
X.~Wang and J.~Yamagishi, ``A comparative study on recent neural spoofing countermeasures for synthetic speech detection,'' \emph{Interspeech 2021}, 2021.

\bibitem{jung2022pushing}
J.-w. Jung, Y.~J. Kim, H.-S. Heo, B.-J. Lee, Y.~Kwon, and J.~S. Chung, ``Pushing the limits of raw waveform speaker recognition,'' \emph{Proc. Interspeech}, 2022.

\bibitem{muppalla2023integrating}
S.~Muppalla, S.~Jia, and S.~Lyu, ``Integrating audio-visual features for multimodal deepfake detection,'' \emph{arXiv:2310.03827}, 2023.

\bibitem{cozzolino2023audio}
D.~Cozzolino, A.~Pianese, M.~Nie{\ss}ner, and L.~Verdoliva, ``Audio-visual person-of-interest deepfake detection,'' in \emph{CVPR}, 2023, pp. 943--952.

\bibitem{abdelnabi2022open}
S.~Abdelnabi, R.~Hasan, and M.~Fritz, ``Open-domain, content-based, multi-modal fact-checking of out-of-context images via online resources,'' in \emph{CVPR}, 2022, pp. 14\,940--14\,949.

\bibitem{huang2023exposing}
M.~Huang, S.~Jia, Z.~Zhou, Y.~Ju, J.~Cai, and S.~Lyu, ``Exposing text-image inconsistency using diffusion models,'' in \emph{ICLR}, 2023.

\end{thebibliography}



\end{document}